\documentclass[conference]{IEEEtran}

\usepackage{cite}
\usepackage{amsmath,amssymb,amsfonts}
\usepackage{algorithmic}
\usepackage{graphicx}
\usepackage{textcomp}
\usepackage{xcolor}
\def\BibTeX{{\rm B\kern-.05em{\sc i\kern-.025em b}\kern-.08em
    T\kern-.1667em\lower.7ex\hbox{E}\kern-.125emX}}

\usepackage{bm}
\usepackage{booktabs}
\usepackage{todonotes}
\usepackage{listings}

\definecolor{dkgreen}{RGB}{0,64,0}
\definecolor{ltgray}{RGB}{245,245,245}
\definecolor{mauve}{RGB}{139,0,139}

\newcommand{\tweakedsim}{\raise.17ex\hbox{$\scriptstyle\mathtt{\sim}$}}

\newcommand{\modelname}{\textit{HPC-Coder}}

\DeclareMathOperator*{\argmax}{arg\,max}

\begin{document}

\title{HPC-Coder: Modeling Parallel Programs using Large Language Models}

\author{\IEEEauthorblockN{Daniel Nichols\IEEEauthorrefmark{2}, Aniruddha Marathe\IEEEauthorrefmark{1}, Harshitha Menon\IEEEauthorrefmark{1}, Todd Gamblin\IEEEauthorrefmark{3}, Abhinav Bhatele\IEEEauthorrefmark{2}}
\IEEEauthorblockA{~\\
\IEEEauthorrefmark{2}\textit{Department of Computer Science, University of Maryland, College Park, MD, USA}\\
\IEEEauthorrefmark{1}\textit{Center for Applied Scientific Computing, Lawrence Livermore National Laboratory, Livermore, CA, USA}\\
\IEEEauthorrefmark{3}\textit{Livermore Computing, Lawrence Livermore National Laboratory, Livermore, CA, USA}\\
Email: dnicho@umd.edu, \{marathe1, gopalakrishn1, tgamblin\}@llnl.gov, bhatele@cs.umd.edu}
}

\maketitle

\begin{abstract}
Parallel programs in high performance computing (HPC) continue to grow in
complexity and scale in the exascale era. The diversity in hardware and
parallel programming models make developing, optimizing, and maintaining
parallel software even more burdensome for developers. One way to alleviate
some of these burdens is with automated development and analysis tools. Such
tools can perform complex and/or remedial tasks for developers that increase
their productivity and decrease the chance for error. Until recently, such
tools for code development and performance analysis have been limited in the
complexity of tasks they can perform, especially for parallel programs.
However, with recent advancements in language modeling, and the availability of
large amounts of open-source code related data, these tools have started to
utilize predictive language models to automate more complex tasks.
In this paper, we show how large language models (LLMs) can be applied to tasks
specific to high performance and scientific codes. We introduce a new dataset
of HPC and scientific codes and use it to fine-tune several pre-trained models.
We compare several pre-trained LLMs on HPC-related tasks and introduce a new
model, \modelname{}, fine-tuned on parallel codes.  In our experiments, we show
that this model can auto-complete HPC functions where generic models cannot,
decorate {\tt for} loops with OpenMP pragmas, and model performance changes in
scientific application repositories as well as programming competition
solutions.

\end{abstract}


\section{Introduction}
\label{sec:introduction}

In recent years, large language models (LLMs) have become the state of the art
for many language modeling related tasks~\cite{zhao2023survey}. Their ability to
model token probabilities within a sequential context make them desirable for
language tasks such as text generation and sequence classification.  In addition
to being used for natural language, such models have recently been applied to
many programming language related
tasks~\cite{codex-copilot-short-author,li2023starcoder,roziere2023code}. The
predictive capabilities of these models translate well to coding tasks, and the
wealth of open-source code available online provides significant data for
training large models.

LLMs trained on source code data have been utilized to automate numerous
software development tasks such as code completion, malware detection, code
refactoring,
etc~\cite{roziere2023code,li2023starcoder,android-malware-detection-ml-survey-2021,ml_for_code,Gu2022AssembleFM,Ahmed2022LearningCS,Haque2022SemanticSM,Ahmad2020ATA,Richter2022CanWL,Kharkar2022LearningTR}.
Additionally, they have been able to automate tasks previously considered
impossible to automate such as code summarization and generation using natural
language. Training LLMs for these tasks requires significant amounts of source
code data that is fortunately available online from open-source code
repositories on GitHub, gitlab etc.  However, this data requirement for training
LLMs is prohibitive for tasks where such data may not exist. One such task is
that of modeling performance (execution time) based on source code. Another 
difficult task is modeling parallel and HPC code where there is less data 
available and it is often more complex code.

Performance data for arbitrary code is difficult to obtain at scale with large
numbers of samples.  First and foremost, it is non-trivial to automate the
collection of performance data for arbitrary source code.  The code needs to be
built and run in order to measure performance, and this process can vary
significantly across repositories.  This can be particularly difficult for
production scientific codes due to code complexity, dependence on external
libraries, and the fact that it often needs to be run in parallel with many
resources.  Second, performance depends on numerous variables besides just the
code such as input problem, architecture, and current machine load/congestion.
These either need to be fixed in the dataset or accounted for within the
modeling pipeline. Finally, source code needs to be considered holistically when
modeling performance, since minor changes in one place may drastically impact
performance elsewhere.  For example, changing the data layout within a data
structure will impact the performance of data access where that structure is
used.  This means that the entirety of the source code needs to be included in
the dataset and performance needs to be collected at a finer granularity.

When a lack of data becomes a hurdle in machine learning tasks, it is typically
solved through data augmentation and/or transfer learning.  Data augmentation
involves extending and/or duplicating data in a manner that still preserves
meaning and representational capacity.  Transfer learning is done by first
training a model on a related or simpler task and then \textit{transferring}
that knowledge to a new problem requiring fewer samples to learn. For our task
we employ transfer learning by using LLMs that have learned to model source code
and then transferring that knowledge to then learn how to model performance of
source code using fewer samples. In particular, we explore modeling parallel
and HPC codes.

In this paper, we utilize LLMs to model high performance and scientific codes,
and then apply that to the problem of performance modeling. In order to
accomplish this, we introduce a new dataset of HPC and scientific codes from
popular open-source repositories. We first demonstrate how our trained model,
\modelname{}, outperforms other LLMs on HPC specific tasks such as code
generation and OpenMP pragma labeling. A set of code generation tests specific
to HPC are introduced and the model can pass these at up to $53\%$ higher
rate than the other models. Additionally, it is able to label {\tt for} loops
with OpenMP pragmas with $97\%$ accuracy. Finally, we demonstrate how the model
can predict relative performance of source code changes with up to $92\%$
accuracy. In summary, this paper makes the following contributions:
\begin{itemize}
    \item A large curated dataset containing HPC and scientific code from numerous
        open-source repositories.
    \item We present an LLM, \modelname{}, fine-tuned to model HPC and scientific 
        code. We show that it trains to better language modeling scores over 
        HPC related code than other state-of-the-art models.
    \item We introduce a set of HPC code generation tasks and demonstrate that
        our model completes these tasks at a significantly better rate than other models on
        HPC-specific code.
    \item We demonstrate how our model can be used to predict OpenMP pragmas 
        with high accuracy.
    \item We utilize our model to predict relative performance of source code 
        changes for two distinct datasets from scientific application repositories and coding competition solutions.
\end{itemize}

\section{Background}
\label{sec:background}
This section provides background on transformer-based language models
and how they can be applied to source code.

\subsection{Large Language Models}
\label{sec:bg_llms}

When applying machine learning to textual data we need a model that
takes text as input and, through the process of training on previous 
data, learns how to predict some property of that text.
In recent years such models have been mostly
dominated by large transformer-based models.
Transformers were first introduced by Vaswani et al.~\cite{transformer}.
They are designed to work with sequential data much like recurrent 
and long short-term memory neural networks.
However, they differ in their use of a self-attention mechanism to 
attribute importance weights to inputs into the model.
Due to this mechanism transformers also process entire sequences
at once unlike recurrent neural networks.

These self-attention units make up the basis of transformer networks.
Weights are divided into query, key, and value weights (namely 
$W_Q$, $W_K$, $W_V$).
These are multiplied by each input token $i$ and stacked to form the 
matrices $Q$, $K$, and $V$, respectively.
Given these matrices and the dimensions of the key vector $d_k$ the 
attention can be computed as below.
$$ \textrm{Attention}\left(Q, K, V\right) = \textrm{softmax}\left( \frac{QK^T}{\sqrt{d_k}} \right) V $$

These weight matrices form a single attention head.
Typically transformers employ several attention heads to form a 
multi-attention head layer.
Having multiple attention heads allows each of them to learn, or 
\textit{attend to}, different abstractions in the input, such as 
parts-of-speech for natural language input.

Generally these networks are trained to model the conditional 
probability of observing a language token or a sequence of tokens. 
For instance, given a string of observed tokens 
$t_1 t_2 \ldots t_{i-1}$ we may want the most likely 
next token $t_i$.
$$ t_i = \argmax_{t} P\left( t_i = t \mid t_1 t_2 \ldots t_{i-1} \right) $$

Similarly we may want to know the probability of a sequence of 
tokens occurring given the entire observed dataset 
$P\left(t_1,t_2,\ldots,t_N\right)$ (i.e. how likely is a given english 
sentence to be real given my previous knowledge of the language).
Using this probability we can define a metric called \textit{perplexity}.
$$ \textrm{Perplexity}(T) = \left( \frac{1}{P(t_1,t_2,\ldots,t_N)} \right)^{\frac{1}{N}} $$

With this metric a model that scores a lower perplexity on its test set
$T$ is better as it assigns a higher probability to the test data.
The ratio is normalized to be invariant to the size of the test set.
Rewriting the formula for perplexity we can see that it is equivalent
to the exponential of the cross-entropy.
\begin{align*}
    \textrm{Perplexity}(T) &= \left( P(t_1,t_2,\ldots,t_N) \right)^{-\frac{1}{N}} \\
    &= \left( \exp \log P(t_1,t_2,\ldots,t_N) \right)^{-\frac{1}{N}} \\
    &= \exp\left( -\frac{1}{N} \log P(t_1,t_2,\ldots,t_N) \right)
\end{align*}

This allows us to train the language model with cross-entropy loss.
Minimizing the loss will, in turn, minimize the perplexity.
The perplexity is recovered by simply taking the exponential of the loss.
It is important to note that perplexity measures model confidence
and not accuracy.
However, it has been demonstrated empirically that lower perplexity generally
leads to better performance on downstream tasks. 

\subsection{Text Generation}
\label{sec:bg_text_generation}

A trained model can then be used to generate new text.
Since the LLM models token probability it may seem simple to select 
the most probable next token, however, this can lead to poor 
text generation.
Often a model's attention puts more focus on on the most recent tokens 
causing this selection method to get stuck in loops or suddenly
forget context.
Most recent works combat this issue by \textit{sampling} from the
model's distribution, but there are several important caveats when
doing this.
For instance, we want to avoid sampling from the tail as this could 
drastically throw off further tokens sampled.
Here we discuss several of the sampling methods used later in this paper
such as temperature, top-$k$, and nucleus sampling.

\vspace{0.06in}
\noindent\textbf{Temperature:}
When sampling temperature controls how \textit{confident} the model 
is in the sampled token.
Lower temperature leads the model to assign more confidence in the 
most likely tokens in the distribution.
On the other end, the model will more uniformly assign confidence 
across the distribution when the temperature is higher.
This term comes from statistical thermodynamics where lower energy 
states are more frequent with a higher temperature.

Temperature is incorporated by dividing the \textit{logits} by the
temperature, $\mathit{temp}$, before computing the softmax output.
The \textit{logits} are the raw, un-normalized outputs of the model 
and the softmax is used to turn this vector into probabilities.
$$ \textrm{softmax}\left( \frac{\bm{logits}}{\mathit{temp}} \right) $$

Thus, as $\mathit{temp} \rightarrow 0$ the output becomes the argmax
and as $\mathit{temp} \rightarrow \infty$ it leads to a uniform sampling.

\vspace{0.06in}
\noindent\textbf{Top-$k$ Sampling:}
In top-$k$ sampling the most likely $k$ tokens are sampled from 
the model.
This aims to exclude the distribution's tail and prevent the model 
from rapidly getting off-topic.
However, this can also reduce the quality of predictions if the 
body of the distribution is wider than $k$.
A common choice for $k$ is 50. 

\vspace{0.06in}
\noindent\textbf{Nucleus Sampling:}
Nucleus, or top-$p$, sampling aims to solve the shortcomings of 
top-$k$ sampling by choosing a more meaningful cut-off point.
In this method the CDF of the distribution is computed and 
sampling is cut-off when the CDF exceeds $p$.
A common choice for $p$ is $0.9$.

\subsection{Using LLMs for Code Generation}
\label{sec:bg_modeling_code}

LLMs can be trained on a variety of downstream tasks and objectives.
When applied to source code data they are typically trained as 
left-to-right, masked, or encoder-decoder models.

\vspace{0.06in}
\noindent\textbf{Left-to-Right:}
Left-to-right or causal language models are trained to predict the 
most probable next token in a sequence.
The model receives and generates text in a left-to-right fashion,
which is where it gets its name.
This limits the amount of context the model can see as it cannot 
use later tokens in its prediction even if they are present in 
the data.
Left-to-right models are useful for text generation related tasks.

\vspace{0.06in}
\noindent\textbf{Masked:}
Unlike left-to-right models, masked models can predict the most 
probable token for any position in the text.
After removing random tokens in the samples and replacing them with 
\textit{mask} tokens, the model is trained to predict the most 
probable tokens to replace the masks with.
In this configuration masked models can make use of more context 
in their predictions.

\vspace{0.06in}
\noindent\textbf{Encoder-Decoder:}
Another common approach is to train a left-to-right model to 
\textit{decode} a sequence after it has been passed through an
encoder.
This type of model can be combined with several different objectives
and is often used with sequence-to-sequence prediction.

\vspace{0.06in}
To apply left-to-right models, which are focused on in this paper, to 
source code you simply need to provide the model with prior context as 
a sequence of tokens and then let it generate new tokens until some stopping 
threshold.
The prior context is typically a natural language comment followed by a 
function declaration.
Tokens are then generated until the function is complete (a closing 
\} bracket in the case of C/C++).

Additionally, when applying language models to code it is typical to 
customize the training process slightly to take advantage of the syntactic 
differences between natural language and code.
For instance, the tokenizer, which is responsible for mapping text to a 
sequence of integers, is often set to group whitespace into single tokens.
This is not necessary in natural language inputs as multiple consecutive 
spaces are uncommon.
However, in code this can meaningfully reduce the sequence size and a formatter
can be applied after code generation to regain formatting.

\section{Overview of the Proposed Methodology}
\label{sec:overview}

\begin{figure}[h]
    \centering
    \includegraphics[width=\columnwidth]{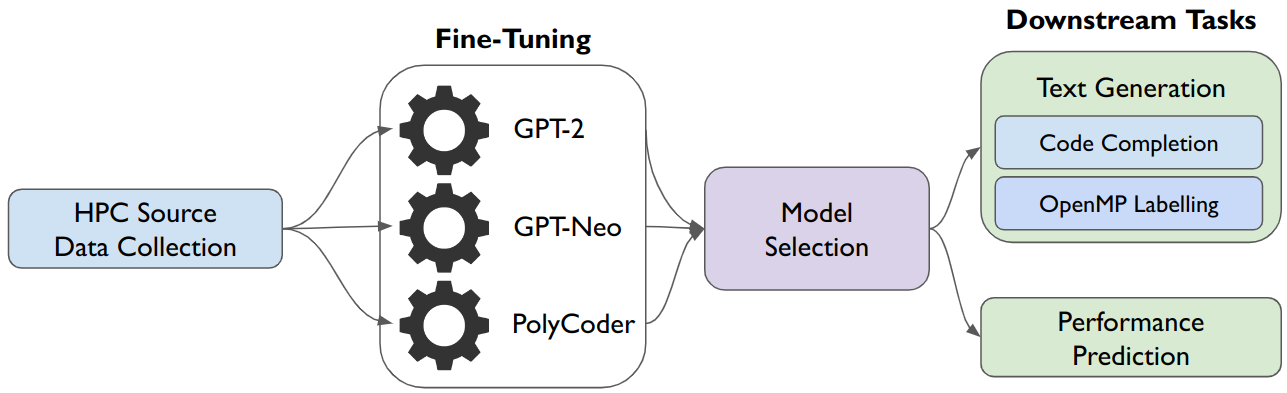}
    \caption{Overview of the steps described in this paper to 
    train an HPC specific model and run it on several downstream tasks.
    After collecting a large dataset of HPC code we fine-tune several pre-trained
    language models and select the best one. The selected model is then used
    to generate code, label OpenMP pragmas, and predict relative performance as 
    part of several downstream tasks.}
    \label{fig:overview}
\end{figure}

Figure~\ref{fig:overview} provides an overview of the data gathering,
training, and downstream application in this paper.
In order to train a large HPC-specific language model we need 
a large dataset of HPC code.
To obtain this, we gather a dataset of HPC source code and use it to fine-tune a 
pre-trained language model.
This data gathering is described in Section~\ref{sec:data-collection} and
presents what HPC sources are used and how they are pre-processed.
Following this, the model fine-tuning and selection are detailed in 
Section~\ref{sec:training} where we explain the training setup and methodology.

We need several realistic tests
to study the performance of the language model on relevant metrics.
We present three main downstream 
tasks for evaluation in Section~\ref{sec:evaluation}.
The first two, code generation and OpenMP pragma labeling, test the model on 
its ability to generate correct and meaningful code.
The last test, relative performance prediction, shows how this trained
model can be used for useful tasks that require language comprehension.
Results from each of these tests are presented and discussed
in Section~\ref{sec:results}.

\section{Data Gathering and Pre-processing}
\label{sec:data-collection}
In order to train a large language model to understand and generate HPC code,
we need to show it lots of examples.
We must first build a dataset to accomplish this.
In this section, we detail our collected dataset and how it is processed.
We present two additional code datasets paired with performance data for
further fine-tuning model performance.

\subsection{HPC Source Code Data}

We first collect a sufficiently large dataset of source code to train the model
on HPC and scientific code.  The HPC source dataset is collected from GitHub
repositories.  The source files are pulled from repositories with C/C++ marked
as the primary language and with $\ge 3$ stars.  The repositories are
additionally filtered by HPC related GitHub \textit{topics}.  Once cloned, we
collect all the C/C++ source files based on their file extension.

\begin{figure}[h]
    \centering
    \includegraphics[width=\linewidth]{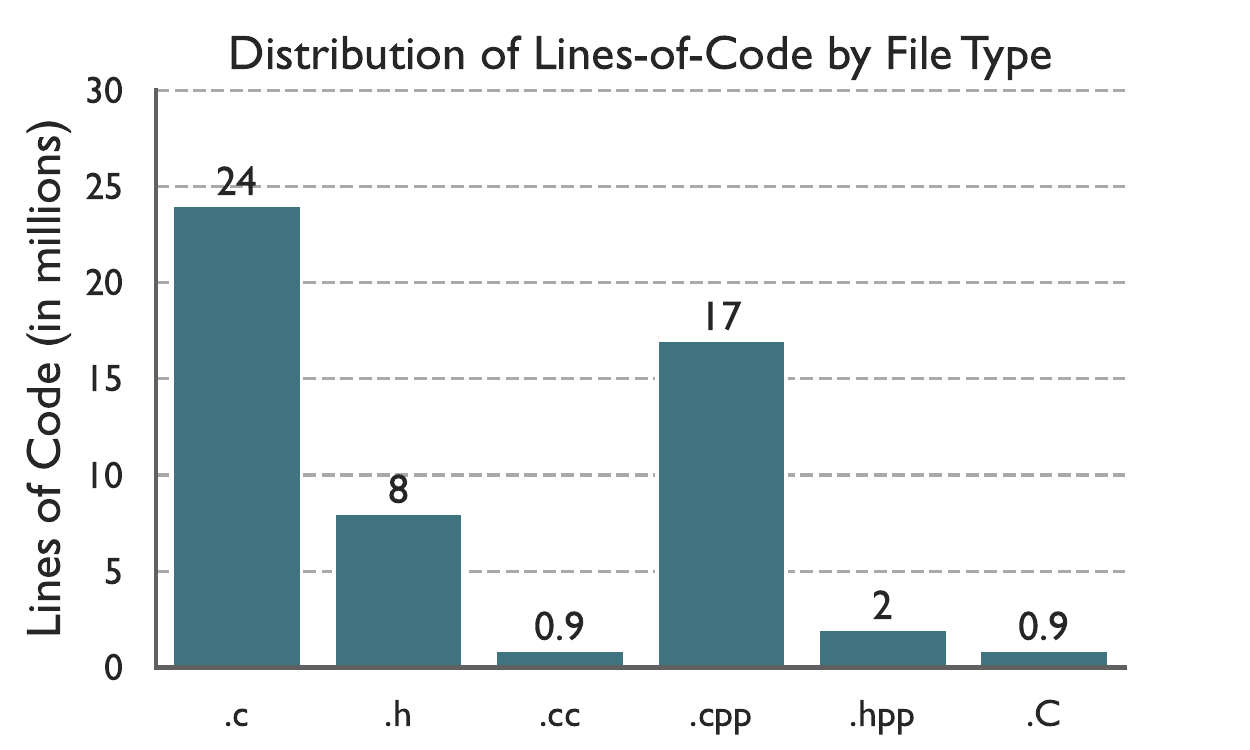}
    \caption{Distribution of no.~of lines of code in each file type. 
        \textit{.cxx}, \textit{.hh}, \textit{.H}, and \textit{.hxx}
        files are included in the dataset, but omitted here due to 
        small counts.}
    \label{fig:dataset-loc-dist}
\end{figure}

This dataset is collected and structured in the same manner as the C/C++ source
dataset from Xu et al.~\cite{xu_2022_code-llms-survey-dataset}.  Their dataset
is scraped from GitHub in a similar manner with the exception of only including
repositories with $\ge 5$ stars.  Figure~\ref{fig:dataset-loc-dist} shows the
distribution of lines of code (LOC) by file types in the HPC source dataset.
There are roughly the same number of LOC in both C and C++ files. The
distribution of actual file counts follows the same trend.

\subsection{Data Pre-processing}

Allamanis~\cite{allamanis_codeduplication} shows how duplicate source 
data, which is prevalent across GitHub repositories, can adversely bias 
LLMs during training.
To prevent this we filter our datasets by removing duplicate files 
based on the hash of their contents.
We use sha256 to hash the contents of the file.

In addition to deduplicating we also filter out small and large files.
Source files larger than 1 MB are designated as large files and removed.
These are generally entire libraries in a single source file or contain 
raw data within the code.
Additionally, files containing less than 15 tokens, as defined by the 
language vocab, are not included.
The reduced dataset sizes after deduplication and filtering are listed 
in Table~\ref{tab:dataset-desc}.
Approximately 18\% of the files are removed during this processing.
Table~\ref{tab:dataset-desc} shows the properties of the dataset 
after each step of deduplication and filtering.

\begin{table}[h]
    \centering
    \caption{Properties of the HPC source code dataset.}
    \label{tab:dataset-desc}
    {\small
    \begin{tabular}{@{}lccc@{}}
    \toprule
    \textbf{Filter}     & \textbf{\# Files} & \textbf{\# LOC} & \textbf{Size (GB)} \\ \midrule
    None                & 239,469           & 61,585,704   & 2.02               \\ 
    Deduplicate         & 198,958           & 53,043,265   & 1.74               \\ 
    \begin{tabular}[c]{@{}c@{}}Deduplicate + remove \\ small/large files\end{tabular} & 196,140           & 50,017,351   & 1.62               \\ \bottomrule
    \end{tabular}
    }
\end{table}

After filtering source files, we tokenize the dataset to obtain integer
values for the text that can be used as input into the model.
We use the pre-trained tokenizers for each of our selected models
(see Section~\ref{sec:training}).
These are all GPT-2~\cite{gpt-2} based Byte-Pair Encoding (BPE) tokenizers.

\subsection{Performance Datasets}\label{sec:perf-datasets}

In addition to the large HPC source code dataset, we create two datasets of
code paired with performance data. These datasets contain code pairs with
performance data for both codes in the pair, and can be used to train an LLM to
model performance characteristics between them.

We create two datasets -- one with pairs of code that are functionally
different and one where they are the same.
The first dataset is created by using version control history to
capture performance regressions. We run each commit for the
Kripke~\cite{kunen2015kripke} and Laghos~\cite{laghos-proxy-app} applications.
These are small HPC apps meant to mimic the computational behavior of larger
scientific applications. We automate building and running each commit to the
best of our ability and collect performance results for 830 commits in total.

The second dataset is a set of programming competition solutions from the {\it
code\_contests} dataset~\cite{alphacode-2022}. These are aggregated from
several online programming competitions: Aizu, AtCoder, CodeChef, CodeForces,
and HackerEarth. This dataset allows us to create pairs of code that solve the
same problem (the contest problem), but may be different in implementation. We
run every correct solution for each problem in the dataset, with the
corresponding problem's test cases as inputs, and record the run time. Using all
the C++ solutions in the dataset we create \tweakedsim 1.7 million samples of
code. Using the run times, we group the solutions into pairs and label
them as {\it slower} and {\it faster} pairs.

\section{Fine-tuning Methodology}
\label{sec:training}
In this section, we describe the models used and how they were selected.
We also discuss the methods used to fine-tune them on our collected dataset.

\subsection{Models Selected For Fine-tuning}

Recent years have seen the introduction of a significant number of large
language models.  These models can range in size from 100 million to more than
100 billion parameters.  Such large models have been shown to work well for
language modeling, but pose significant hurdles to train and use in practice.
They can take months to train on large GPU clusters and typically cannot
feasibly run inference on consumer-grade hardware.  Thus, choosing the right
model requires selecting one that can sufficiently model the language data, but
also be reasonably deployed for downstream tasks.

\setlength{\tabcolsep}{3pt}
\begin{table}[h]
    \centering
    \caption{Description of the models used for fine-tuning.}
    \label{tab:model-descriptions}
    {\footnotesize
    \begin{tabular}{@{}lccccc@{}}
    \toprule
    \textbf{Model}                                             & \textbf{\# Params.} & \textbf{\# Layers} & \textbf{\begin{tabular}[c]{@{}c@{}}Hidden \\ Size\end{tabular}} & \textbf{\begin{tabular}[c]{@{}c@{}}Window\\ Size\end{tabular}} & \textbf{\begin{tabular}[c]{@{}c@{}}Pre-Training\\ Set\end{tabular}} \\ \midrule
    \textbf{GPT-2}~\cite{gpt-2}                                & 1.5B              & 48              & 1600                                                            & 1024                                                           & WebText~\cite{openwebtext}                                          \\
    \textbf{GPT-Neo}~\cite{gpt-neo}                            & 2.7B              & 32              & 2560                                                            & 256                                                            & Pile~\cite{the_pile}                                                \\
    \textbf{PolyCoder}~\cite{xu_2022_code-llms-survey-dataset} & 2.7B              & 32              & 2560                                                            & 2048                                                           & Source Code                                                         \\ \bottomrule
    \end{tabular}
    }
\end{table}

Keeping the above mentioned requirements in mind, we select several models for
fine-tuning and/or testing.  These are listed in
Table~\ref{tab:model-descriptions}.  All of these are based on
GPT-2~\cite{gpt-2} and/or GPT-3~\cite{gpt-3} architectures with slight
variations in size, configuration, and pre-training data.  GPT-2, the smallest
in our experiments, is pre-trained on the WebText~\cite{openwebtext} dataset,
which is a collection of language data scraped from the internet.  We use the
1.5 billion parameter GPT-2 model variant in this paper.
PolyCoder~\cite{xu_2022_code-llms-survey-dataset} is pre-trained on a
collection of solely source code data from GitHub that contains a mixture of 12
popular programming languages~\cite{xu_2022_code-llms-survey-dataset}.  Between
these two is GPT-Neo~\cite{gpt-neo} that is pre-trained on the Pile
dataset~\cite{the_pile}.  This dataset contains a collection of approximately
800GB of text data from the internet, academic articles, source code, etc.
Notably this dataset has a mixture of natural language and code.  It has
been demonstrated that pre-training over \textit{both} natural language and
code can improve the performance of the model.

We exclude models such as GPT-4~\cite{openai2023gpt4}, the
state-of-the-art model that powers GitHub CoPilot, from our experiments due to
the model and its dataset being closed source. It is currently only accessible
for inference via a non-free API. GPT-4's dataset being closed source is
significant as we cannot remove data it has trained on from the dataset we use
to evaluate its performance, so its results would be overly optimistic. This
prevents a realistic evaluation and comparison.

\subsection{Fine-tuning Setup and Hyperparameters}

We rely on the functionality provided in the HuggingFace~\cite{huggingface}
Python library for fine-tuning the models.  This library automates many of the
tasks related to loading and pre-processing datasets, and running language
models on the datasets.  In particular, we use the \verb|Trainer| interface
with DeepSpeed~\cite{deepspeed-extreme-3d} as the backend to optimize
fine-tuning.  DeepSpeed is a framework that provides distributed training
functionality and several memory optimizations to enable large models to fit in
GPU memory.

Starting with the pre-trained models, we fine-tune them on a single node with
an AMD EPYC 7763 CPU, 512 GB memory, and four 40 GB NVIDIA A100 GPUs.
With DeepSpeed's ZeRO memory optimizations~\cite{zero_offload}, all of the
models fit entirely within a single A100 GPU and are, thus, fine-tuned using pure
data parallelism.  We refer the reader to~\cite{bennun2019demystifying,
nichols:arxiv2022} for a comprehensive overview of training deep neural
networks in parallel.

We use the AdamW~\cite{adamw} optimizer for all the models to update model
weights and minimize the loss.  We set the learning rate to $5\times10^{-5}$
and Adam parameters $\beta_1$ and $\beta_2$ to $0.9$ and $0.999$, respectively.
These hyperparameters are consistent with typical values in the literature.
16-bit floating point precision is used to accelerate fine-tuning and reduce
model size on the A100s. We record the perplexity of the
model on the training data during fine-tuning. This is calculated as the exponential of the
training loss (see Section~\ref{sec:bg_llms}). Every 1000 optimizer steps, we
also test the model using the validation dataset, and record the perplexity and
accuracy at predicting tokens. The validation dataset is 5\% of the full
dataset, separate from the training dataset.

\section{Downstream Inference Tasks and Evaluation Metrics}
\label{sec:evaluation}
In this section, we introduce the benchmarks and metrics used to evaluate the
performance of the language models.

\subsection{Code Completion}
\label{sec:code-completion}

A standard benchmark for code generation tasks is the HumanEval
benchmark~\cite{codex-copilot-all-author}. This is comprised of 164 sample
Python problems, where the input to the model is a natural language description
of a function and function header. The model generates code for the function
implementation, and is scored on functional correctness rather than textual
similarity or equivalence.

We introduce our own adaptation of this benchmark for HPC C/C++ programs. Our
benchmark consists of 25 custom HPC code generation problems including simple
numerics, OpenMP parallel code, and MPI routines. Table~\ref{tab:hpc-tests}
lists the tests used in our evaluation. Figure~\ref{code:openmp_saxpy} shows a
sample prompt (top) and output (bottom) for a shared-memory parallel
implementation of \texttt{saxpy}. The prompt is provided as input to the model
and it is expected to generate text functionally equivalent to the text on the
bottom.

\begin{table}[h]
        \centering
        \caption{Code generation tests. OpenMP and MPI columns denote if the test includes a version with that parallel backend.}
        \label{tab:hpc-tests}
        \begin{tabular}{@{}llccc@{}}
        \toprule
        \textbf{Name}           & \textbf{Description}                                                               & \textbf{Seq.} & \textbf{OpenMP} & \textbf{MPI} \\ \midrule
        \textit{Average}        & \begin{tabular}[c]{@{}l@{}}Average of an array\\ of doubles\end{tabular}           & \checkmark          & \checkmark      & \checkmark   \\
        \textit{Reduce}         & \begin{tabular}[c]{@{}l@{}}Reduce by generic\\ function foo\end{tabular}           & \checkmark          & \checkmark      & \checkmark   \\
        \textit{Saxpy}          & Saxpy                                                                              & \checkmark          & \checkmark      & \checkmark   \\
        \textit{Daxpy}          & Daxpy                                                                              & \checkmark          & \checkmark      & \checkmark   \\
        \textit{Matmul}         & \begin{tabular}[c]{@{}l@{}}Double-precision\\ matrix multiply\end{tabular}         & \checkmark          & \checkmark      & \checkmark   \\
        \textit{Simple Send}    & Send MPI message                                                                   &                     &                 & \checkmark   \\
        \textit{Simple Receive} & Receive MPI message                                                                &                     &                 & \checkmark   \\
        \textit{FFT}            & Double-precision FFT                                                               & \checkmark          & \checkmark      & \checkmark   \\
        \textit{Cholesky}       & \begin{tabular}[c]{@{}l@{}}Single-precision Cholesky\\ factorization\end{tabular} & \checkmark          & \checkmark      & \checkmark   \\
        \textit{Ping-pong}      & MPI ping-pong                                                                      &                     &                 & \checkmark   \\
        \textit{Ring pass}      & MPI ring pass                                                                      &                     &                 & \checkmark   \\ \bottomrule
        \end{tabular}
\end{table}

\begin{figure}[!ht]
\centering    
(a) Prompt
\begin{lstlisting}
/*
 multiply scalar float a by vector x and add to y
 vectors x and y are length N
 use OpenMP to compute in parallel
*/
void saxpy(float *x, float *y, float a, int N) {
\end{lstlisting}
\vspace{0.1cm}
(b) Output
\begin{lstlisting}
    #pragma omp parallel for
    for (int i = 0; i < N; i++) {
        y[i] += a * x[i];
    }
}
\end{lstlisting}
\caption{An example prompt asking the model to generate a parallel
version of saxpy. The comment and function header make up the prompt.
The function body on the bottom shows a potential model output.}
\label{code:openmp_saxpy}
\end{figure}

\vspace{0.08in}
\noindent\textbf{Evaluation Metric:} We first record the ratio of generated
samples that build correctly to those that do not. This indicates the model's
ability to generate syntactically correct code. For those that compile we
compute the $\textrm{pass}@k$ metric that denotes the probability that at least
one of k samples out of $N_p$ code samples is correct.  We do $N_p$ trials with
each prompt $p$ to generate $N_p$ code samples, compile/run the samples, and
record the number that are functionally correct ($c_p$).  To estimate the
probability that at least one of $k$ samples chosen from $N_p$ samples is
correct for a particular prompt, $p$, we can use the number of generated samples
that are functionally correct, $c_p$, out of the $N_p$ total samples generated
to calculate $\textrm{pass}@k$ for a given $k$ as,
\begin{equation}
\textrm{pass}@k = 1 - \binom{N_p - c_p}{k} / \binom{N_p}{k}
\end{equation}
For each model, we report the $\textrm{average\_pass}@k$ metric as the average
$\textrm{pass}@k$ over all $P$ prompts as shown below:
\begin{equation}
\label{eq:pass_k}
    \textrm{average\_pass}@k = \frac{1}{P} \sum_{i=1}^{P} \left[ 1 - \frac{\binom{N_i - c_i}{k}}{\binom{N_i}{k}} \right]
\end{equation}

This metric provides insight into the probability of a model generating
functionally correct code. In our experiments, we calculate the
$\textrm{pass}@k$ score for several temperatures, namely 0.1, 0.2, 0.4, 0.6, and
0.8, and select the best one. This is in line with experiments in related
literature~\cite{xu_2022_code-llms-survey-dataset}. For each temperature and
prompt, we generate $N_p = 100$ samples. The code is generated with nucleus
sampling using $0.93$ as the cutoff value in the CDF (see
Section~\ref{sec:background}).

To compile the generated code samples, we use \texttt{g++} with the
``\texttt{-O2} \texttt{-std=c++17} \texttt{-fopenmp}" flags. For tests that need
MPI we use the OpenMPI \texttt{mpicxx} compiler. If the build is successful,
then a corresponding driver binary is called that will call and test the
generated function for correctness. These are run on a AMD EPYC 7763 CPUs with
64 physical cores at 2.45 GHz each. For tests that require OpenMP or MPI we only
denote them as correct if they used the corresponding parallel framework to
compute their result.

\subsection{Predicting OpenMP Pragmas}
\label{sec:omp-pragmas}

A common HPC coding task is decorating \verb|for| loops with OpenMP pragmas.
Every pragma starts with \verb|#pragma omp parallel for| and is followed by a
list of optional clauses that modify the behavior of the parallel \verb|for|.
We test the model's ability to
write OpenMP pragmas for arbitrary \verb|for| loops.

\vspace{0.08in}
\noindent\textbf{Further Fine-tuning:} 
We cannot directly use the existing models to generate pragmas {\it before} a
\verb|for| loop, since they are all left-to-right and can only append tokens to
sequences. Thus, we need to further fine-tune the models on a smaller dataset
that puts the \verb|for| loop before the pragma. To accomplish this, we first
create a dataset of every \verb|for| loop with an OpenMP pragma from our HPC
code dataset. 500 tokens of context from before the \verb|for| loop are also
included. This results in a dataset with 13,900 samples.

Since our model is left-to-right, we format each sample by moving the pragma to
directly after the loop and a unique separating token
\verb|<begin-omp>|. This allows us to use the model by providing a \verb|for|
loop plus some context and the model will generate an OpenMP pragma for
the \verb|for| loop.

Each model is fine-tuned on this smaller dataset for three epochs (passes over the
entire dataset). To prevent overfitting we use a starting learning rate of
$3\times10^{-5}$. During training 10\% of the dataset is set aside for
validation.

\vspace{0.08in}
\noindent\textbf{Evaluation Metric:}
To measure the success of this test, we use the accuracy of generating correct
pragmas. This is calculated as shown in Equation~\ref{eq:accuracy}.
\begin{equation}\label{eq:accuracy}
    \textrm{accuracy} = \frac{\textrm{\# correct pragmas}}{\textrm{total pragmas tested}}
\end{equation}
For this problem, we define a \textit{correct} pragma in two ways: syntactic and
functional. To measure syntactic correctness we compare the generated pragma
with the actual pragma for textual equivalence. Since it is impossible to
automate the running and evaluation of arbitrary \verb|for| loops from our
dataset we measure functional correctness by comparing the generated pragmas
with the actual ones while ignoring differences that do not contribute to
functionality. For instance we ignore reordering of variables and clauses where
these do not matter. Additionally, clauses such as \textit{schedule} are
ignored. This correctness check is done using a custom Python script that parses
the pragmas and compares them. We record accuracy from both of these correctness
metrics for each model.

\subsection{Relative Performance Prediction}
\label{sec:rel-perf-prediction}

In addition to text generation, we can also use the LLMs for classification. Here
we use them to predict performance slowdowns between two pairs of code.

\vspace{0.08in}
\noindent\textbf{Further Fine-tuning:} In order to use the models for relative
performance classification we need to first fine-tune them on new data for this
output task. Using the Git commit data from Section~\ref{sec:perf-datasets} we give the model text
for a region of code before and after a Git commit. The codes are concatenated
with a unique token separating them, namely \verb|<COMMIT>|. We repeat a similar
process for the code contest dataset, but instead separate pairs by the token
\verb|<PAIR>|. With this data the model is fine-tuned to predict whether the
second code will be slower (\textit{positive}) or the same/faster
(\textit{negative}). 

For each dataset we fine-tune the model on 90\% of the data with the other
10\% set aside for evaluation. The model takes the concatenated sequences of the
two versions of the code implementation and is fine-tuned for the binary
classification problem of predicting relative performance. The training
objective is classification accuracy, which we also use to measure success for
this task.

\vspace{0.08in}
\noindent\textbf{Evaluation Metric:} To evaluate the performance on this task
we measure the model's classification accuracy. This is calculated as shown in
Equation~\ref{eq:class-accuracy}.
\begin{equation}\label{eq:class-accuracy}
        \textrm{accuracy} = \frac{\textrm{\# correct performance predictions}}{\textrm{total performance predictions}}
\end{equation}
For this metric higher is better and a classification accuracy of 100\%
signifies a perfect score.

\section{Results}
\label{sec:results}
We now present the fine-tuning and evaluation results using the
downstream tasks discussed in Section~\ref{sec:evaluation}.

\subsection{Fine-tuning on HPC Source Code Data}

We first show the results of fine-tuning the three models selected in
Table~\ref{tab:model-descriptions}. Table~\ref{tab:perplexities} shows the
validation perplexity at the end of fine-tuning. Here perplexity is
calculated as the exponential of the loss as described in
Section~\ref{sec:background}. Each model converges to a low perplexity score
over the separate testing set (between 2 and 4). GPT-Neo and PolyCoder achieve
comparable perplexity scores (within 0.01) while GPT2 achieves a higher
perplexity. All three have different pre-training datasets and the former two
are of a larger size than GPT2 (see Table~\ref{tab:model-descriptions}). From
this we can conclude that for this problem the pre-training dataset had less of
an impact on validation perplexity than the model size. The lower perplexity of
the larger models means that they model the language better.

\begin{table}[h]
    \centering
    \caption{Final validation perplexities for each model after fine-tuning on the HPC source code dataset.}
    \label{tab:perplexities}
    \begin{tabular}{lccc}
    \toprule
    \textbf{Model} & GPT-2 & GPT-Neo & PolyCoder \\ \midrule
    \textbf{Final Validation Perplexity} & 4.47 & 2.23 & 2.24 \\ \bottomrule
    \end{tabular}
\end{table}

For the rest of the results presented in this section we will use PolyCoder+HPC,
GPT-Neo+HPC, and GPT2+HPC to refer to the respective models fine-tuned
on the HPC dataset.

After fine-tuning each of the models and evaluating them on the downstream tasks
we noticed that the perplexity would keep improving with more fine-tuning, but
the downstream evaluation performance would start to decrease. This is likely
because LLMs are subject to \textit{catastrophic forgetting} during fine-tuning.
\textit{Catastrophic forgetting} is the phenomenon where previously learned
information is lost or forgotten as the model continues training and updating
its weights. It is typically prevented by minimizing the amount of fine-tuning
and using a sufficiently low learning rate.

\begin{figure}[h]
    \centering
    \includegraphics[width=\columnwidth]{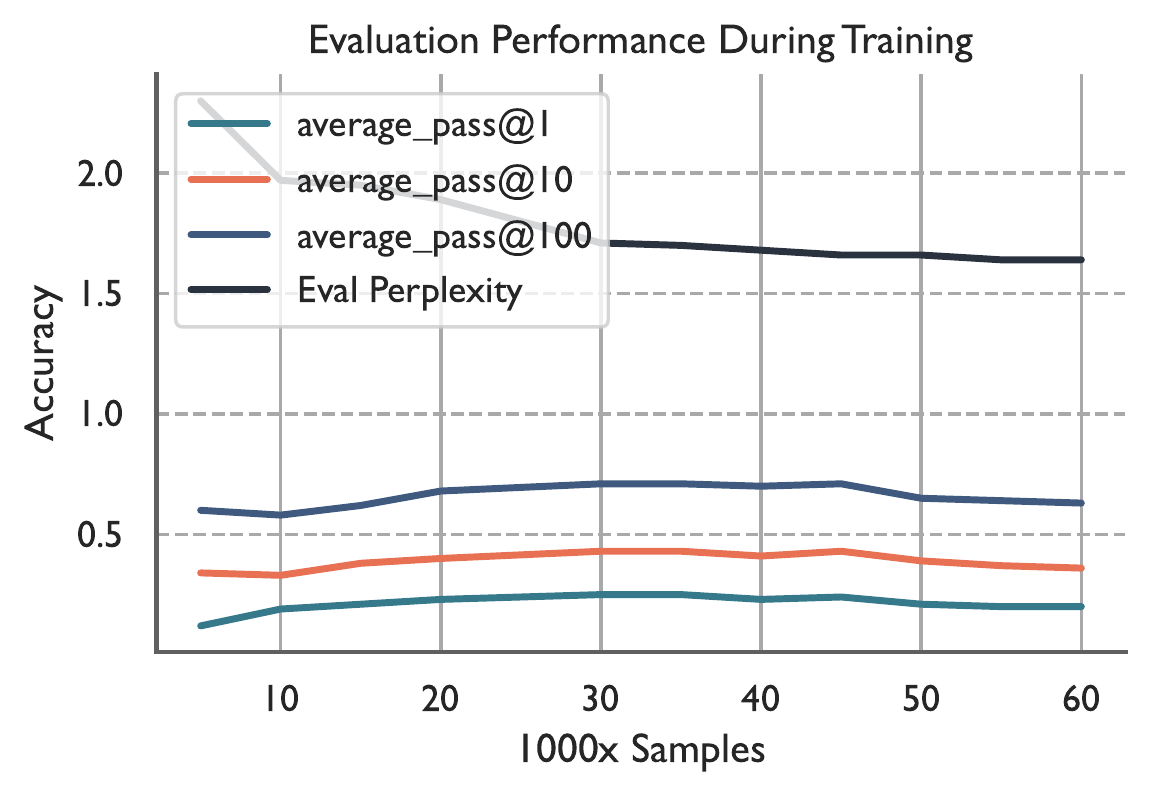}
    \caption{Downstream evaluation performance across training iterations
        for PolyCoder+HPC.
        The model starts to perform worse around 45,000 samples even though
        the perplexity keeps improving.}
    \label{fig:tuning-results}
\end{figure}

To explore this phenomenon we ran the code generation tasks every 1000 samples
during fine-tuning of the PolyCoder model. Figure~\ref{fig:tuning-results}
presents the results from our evaluation tests during fine-tuning on the PolyCoder
model. After seeing about 45,000 samples during fine-tuning the model starts to
decrease in evaluation performance. This is in contrast to the perplexity which
keeps improving past 45,000 samples. Based on this result we stop fine-tuning at
45,000 samples and use these weights for the rest of the evaluations.
Additionally, due to the computation time needed to run this test we use the
45,000 samples stopping point for fine-tuning all the models.

\subsection{Code Completion}

\begin{figure}[h]
    \centering
    \includegraphics[width=\columnwidth]{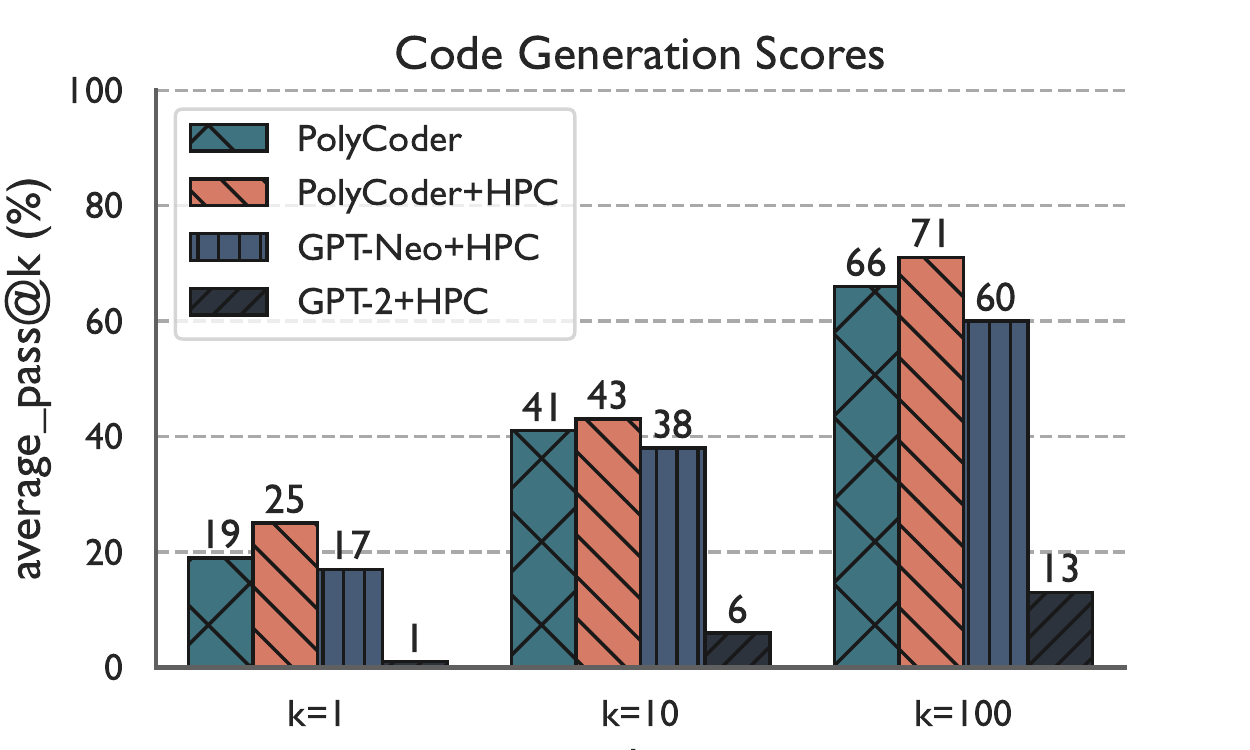}
    \caption{Comparison of models on code generation. The clusters
    represent the average pass@k scores for $k=1,10$ and $100$. Higher percentage is 
    better.}
    \label{fig:code-generation-results}
\end{figure}

\begin{figure}[h]
    \centering
    \includegraphics[width=\columnwidth]{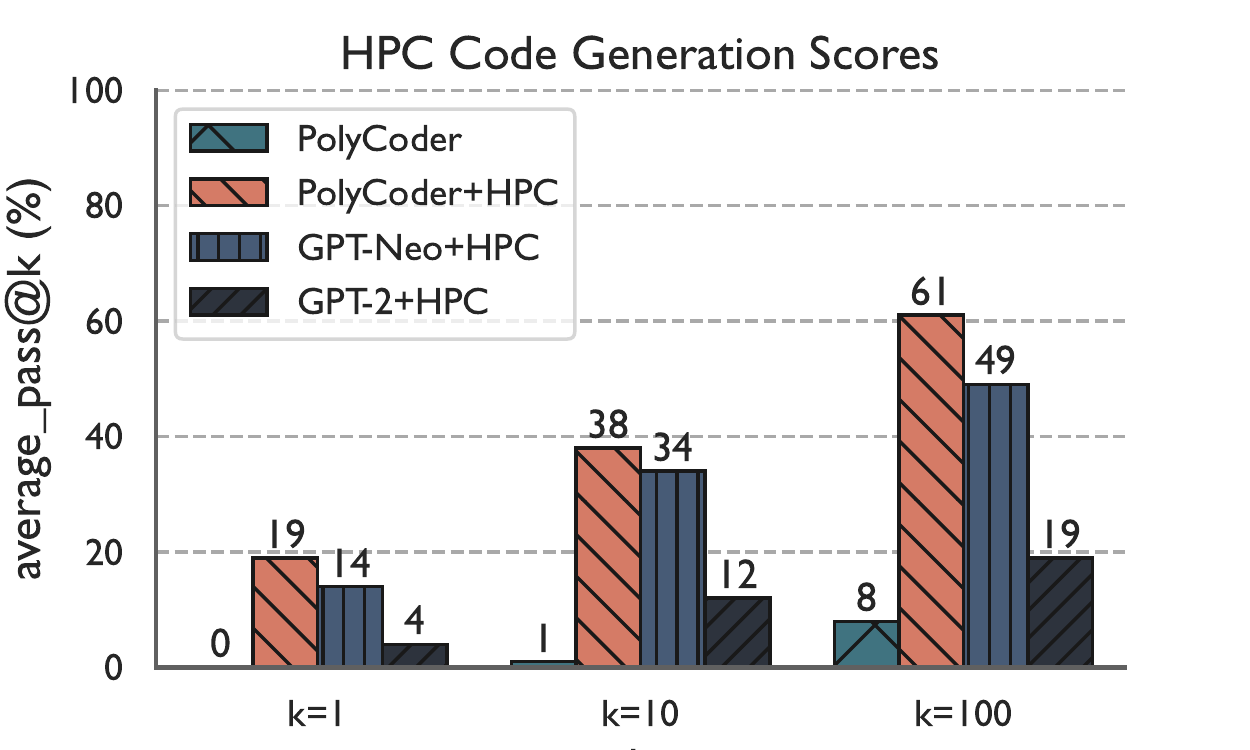}
    \caption{Comparison of models on code generation for HPC-specific
    functions. The clusters
    represent the average pass@k scores for $k=1,10$ and $100$. Higher percentage
    is better.}
    \label{fig:hpc-code-generation-results}
\end{figure}

Having fine-tuned the three models, we now start using them for the different downstream tasks described in Section~\ref{sec:evaluation}. The first downstream task is code generation,
described in
Section~\ref{sec:code-completion}. Figure~\ref{fig:code-generation-results}
shows the $\textrm{average\_pass@k}$ rates for the code generation tests. The $\textrm{average\_pass@k}$ values
are computed according to Equation~\ref{eq:pass_k}. We use PolyCoder as a
baseline for comparison since it is a state-of-the-art LLM for code generation.
PolyCoder+HPC scores the best for average pass@1, pass@10, and pass@100. For
each value of $k$ the models score in the order of PolyCoder+HPC, PolyCoder,
GPT-Neo+HPC, and GPT2+HPC. PolyCoder+HPC gains the slight edge over the original
PolyCoder by successfully generating code for the HPC-specific tasks (see
Figure~\ref{fig:hpc-code-generation-results}).

In Figure~\ref{fig:code-generation-results} we see that GPT2+HPC scores
significantly lower than the other models. This is likely due to the smaller
model size and the fact that there is no source code in its pre-training
dataset. In this instance fine-tuning is not enough to enable GPT-2 to generate
correct C++ HPC code.

Altogether, the scores are indicative that PolyCoder+HPC and GPT-Neo+HPC has
learned how to generate valid C++ code. For instance, if the best model,
PolyCoder+HPC, is permitted to generate 100 samples, then 71\% of them are
correct on average across all the tests. Similarly for 1 sample generated this
is 25\%. These numbers roughly align with results
from~\cite{xu_2022_code-llms-survey-dataset} on the HumanEval Python tests.
However, the results are not directly comparable since they are a different set
of tests in a different programming language.

To demonstrate the generative capabilities of the specialized models we reduce
the code generation tasks to those that are specific to HPC. This includes code
that uses OpenMP and/or MPI parallelism.
Figure~\ref{fig:hpc-code-generation-results} shows the performance when
restricted to these tests. We see that PolyCoder is unable to generate OpenMP
and MPI code as it scores significantly lower than the rest. GPT2+HPC still
performs fairly low, however, its score has actually improved slightly over
Figure~\ref{fig:code-generation-results}. This is due to the fact that it has
only seen HPC-specific code during training and that is what is
being tested here. 

\begin{figure}[h]
    \centering
    \includegraphics[width=\columnwidth]{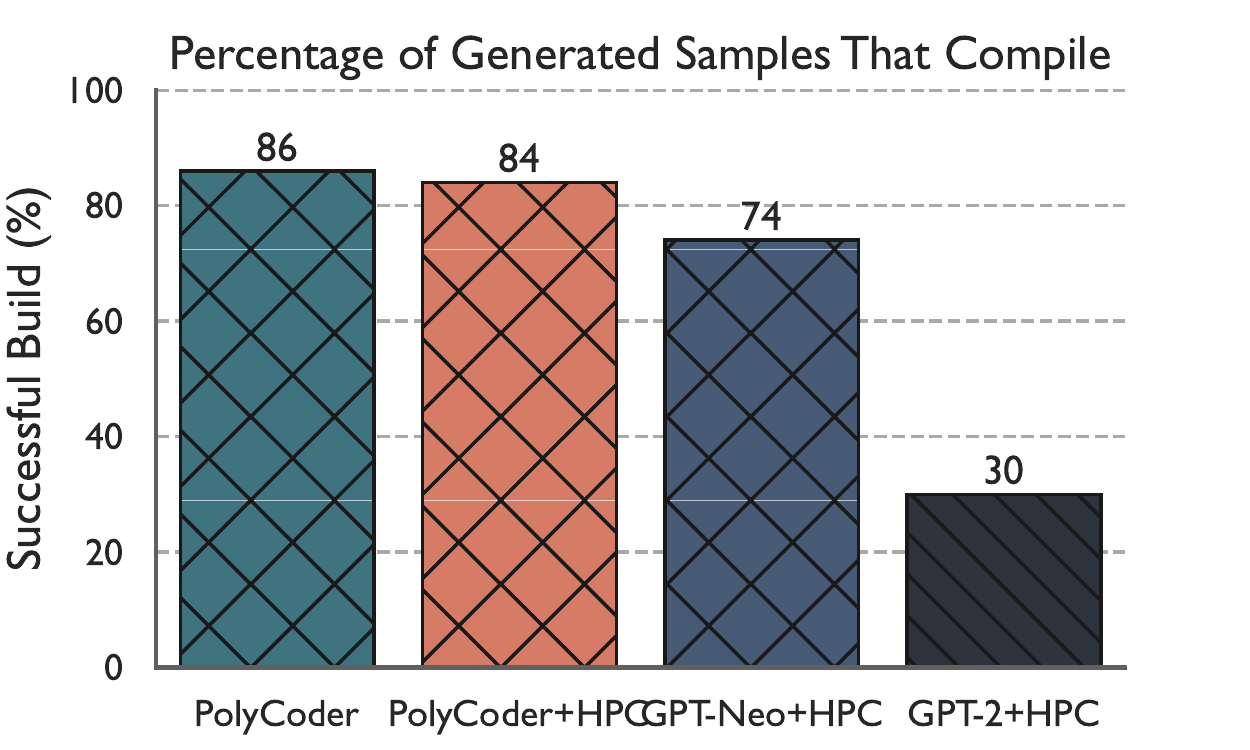}
    \caption{Comparison of the models' build rate. Both PolyCoder and 
    PolyCoder+HPC have the best percentage of total samples that successfully
    compile. Higher percentage is better.}
    \label{fig:build-rate}
\end{figure}

Another point of interest besides functional correctness is syntactic
correctness. This can be measured by the total number of generated samples that
compile successfully. This is how often the model generates valid code, whether
it is functionally correct or not. This data is presented in
Figure~\ref{fig:build-rate}. PolyCoder and PolyCoder+HPC both perform the best
compared to the other models with 84\% and 86\% of samples compiling correctly,
respectively. GPT-Neo+HPC performs slightly worse at 74\% and GPT2-HPC has only
30\% of samples compile. The worse performance of the latter two can likely be
attribute to their pre-training datasets having less code. We also
observe that for all models there is a visual correlation between build and
correctness rates, which is expected as a model needs to compile in order to
be functionally correct.

The code in Figure~\ref{code:openmp_sum} shows example output from PolyCoder and
PolyCoder+HPC on generating OpenMP code to compute a sum in parallel. We see
that PolyCoder is able to produce correct sequential code, however, it fails to
add any OpenMP pragmas. The PolyCoder+HPC model is able to correctly tag the for
loop with an OpenMP pragma.

\begin{figure}[!ht]
\centering
(a) Prompt
\begin{lstlisting}
/*
 Compute the sum of the array X and return the sum. 
 X has N elements. 
 Use OpenMP to compute the sum in parallel.
*/
float sum(float *X, int N) {
\end{lstlisting}
\vspace{0.1cm}
(b) PolyCoder output
\begin{lstlisting}
    float sum = 0.0f;
    for (int i = 0; i < N; i++)
        sum += X[i];
    return sum;
}
\end{lstlisting}
\vspace{0.1cm}
(c) PolyCoder+HPC output
\begin{lstlisting}
    float sum = 0.0f;
    #pragma omp parallel for reduction(+:sum)
    for (int i = 0; i < N; i++)
        sum += X[i];
    return sum;
}
\end{lstlisting}
\caption{Example OpenMP output from (b)~PolyCoder and (c)~PolyCoder+HPC. The comment and function description (top) make up the prompt that 
is given to the model, while the bottom two blocks are the generated
text. We see that PolyCoder is unable to generate OpenMP pragmas for the 
reduction in this example.}\label{code:openmp_sum}
\end{figure}

A similar example with distributed memory is shown in
Figure~\ref{code:mpi_average}. The PolyCoder+HPC is able to generate correct MPI
code to compute an average in parallel across ranks. PolyCoder often generated
long and incorrect code that demonstrated little understanding of how to write
MPI routines. The example presented is indicative of the rest of the MPI samples
generated by PolyCoder. Notably, some instances did contain MPI calls, so
PolyCoder must have some MPI code in its pre-training dataset. However, it
lacked the ability to properly use MPI.

\begin{figure}[!ht]
\centering
(a) Prompt
\begin{lstlisting}
/*
    Use MPI to compute the average of X across ranks
    and return the result on all ranks. 
    Use the default communicator.
*/
double mpiAverage(double X) {
\end{lstlisting}
\vspace{0.1cm}
(b) PolyCoder output
\begin{lstlisting}
    int rank;
    int nprocs;
    double average;
    MPI_Status status;
    MPI_Request request;

    MPI_Init(&argc, &argv);
    MPI_Comm_rank(MPI_COMM_WORLD, &rank);
    MPI_Comm_size(MPI_COMM_WORLD, &nprocs);

    if (rank == 0) {
        MPI_Recv(&status, 1, MPI_INT, 0, 0, MPI_COMM_WORLD,
                 &request);
        MPI_Send(&X, 1, MPI_INT, 0, 0, MPI_COMM_WORLD);
        MPI_Send(&average, 1, MPI_DOUBLE, 0, 0,
                 MPI_COMM_WORLD);
    }
    return average;
}
\end{lstlisting}
\vspace{0.1cm}
(c) PolyCoder+HPC output
\begin{lstlisting}
    int size;
    double sum;

    MPI_Comm_size(MPI_COMM_WORLD, &size);
    MPI_Allreduce(&X, &sum, 1, MPI_DOUBLE, MPI_SUM, 
                  MPI_COMM_WORLD);

    return sum / size;
}
\end{lstlisting}
\caption{Example MPI output from (b)~PolyCoder and (c)~PolyCoder+HPC. 
The highlighted region is code
generated by the model (reformatted to fit the column). PolyCoder results 
varied significantly, however, the above example demonstrates the general lack 
of understanding it had for MPI.}\label{code:mpi_average}
\end{figure}

\begin{figure}[h]
    \centering
    \includegraphics[width=\linewidth]{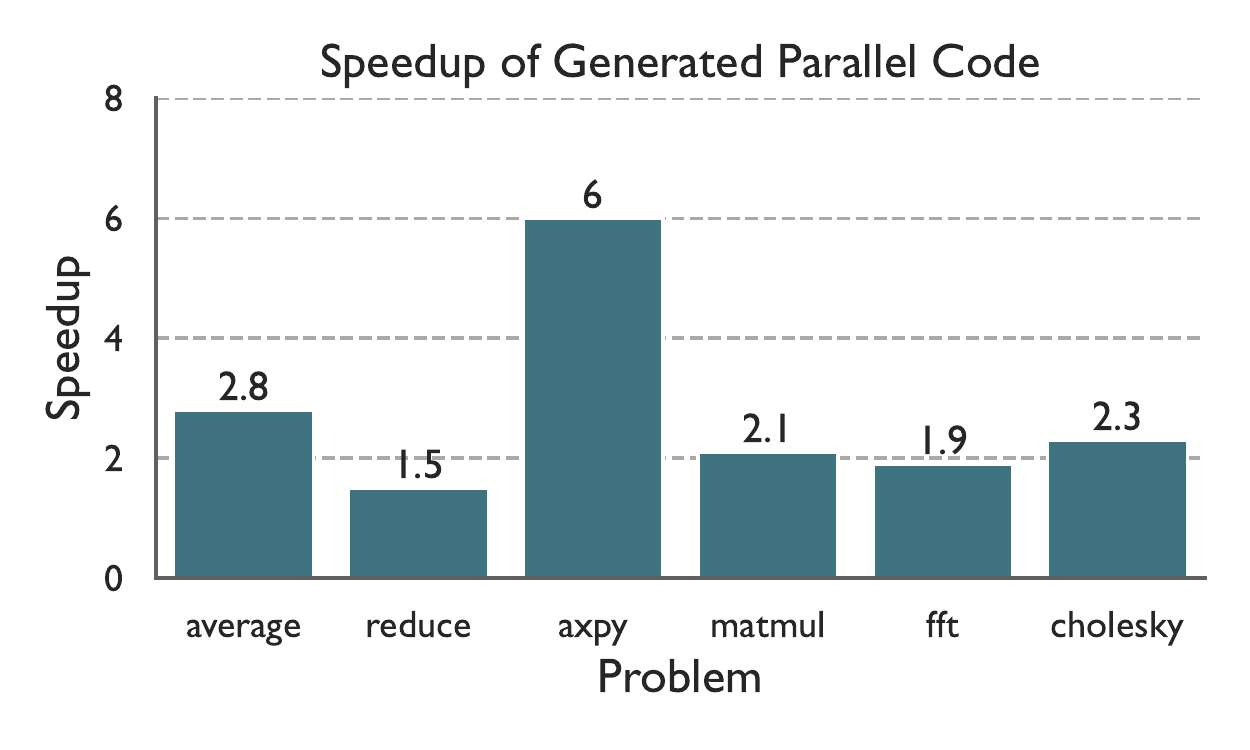}
    \vspace{-0.3in}
    \caption{Comparison of the speedups for the code generation tests over 
        sequential baselines. They are all above 1 demonstrating that 
        the model is not generating very poor performing parallel code.}\label{fig:run-times}
\end{figure}

Figure~\ref{fig:run-times} shows the speedups for the code generated by
PolyCoder+HPC over sequential baselines. These are hand-written efficient,
sequential implementations for each test. We see that PolyCoder+HPC is able to
generate code that is faster than the sequential baseline. This demonstrates
that it is not generating very poor performing parallel code and is likely using
the parallelism correctly.

Since PolyCoder+HPC scores the highest in training and these code generation
tests we select it for further comparisons in the rest of the paper.
PolyCoder+HPC is the fine-tuned model we present as HPC-Coder. We continue to
use PolyCoder as a baseline.

\subsection{Predicting OpenMP Pragmas}

\begin{figure}[h]
    \centering
    \includegraphics[width=\columnwidth]{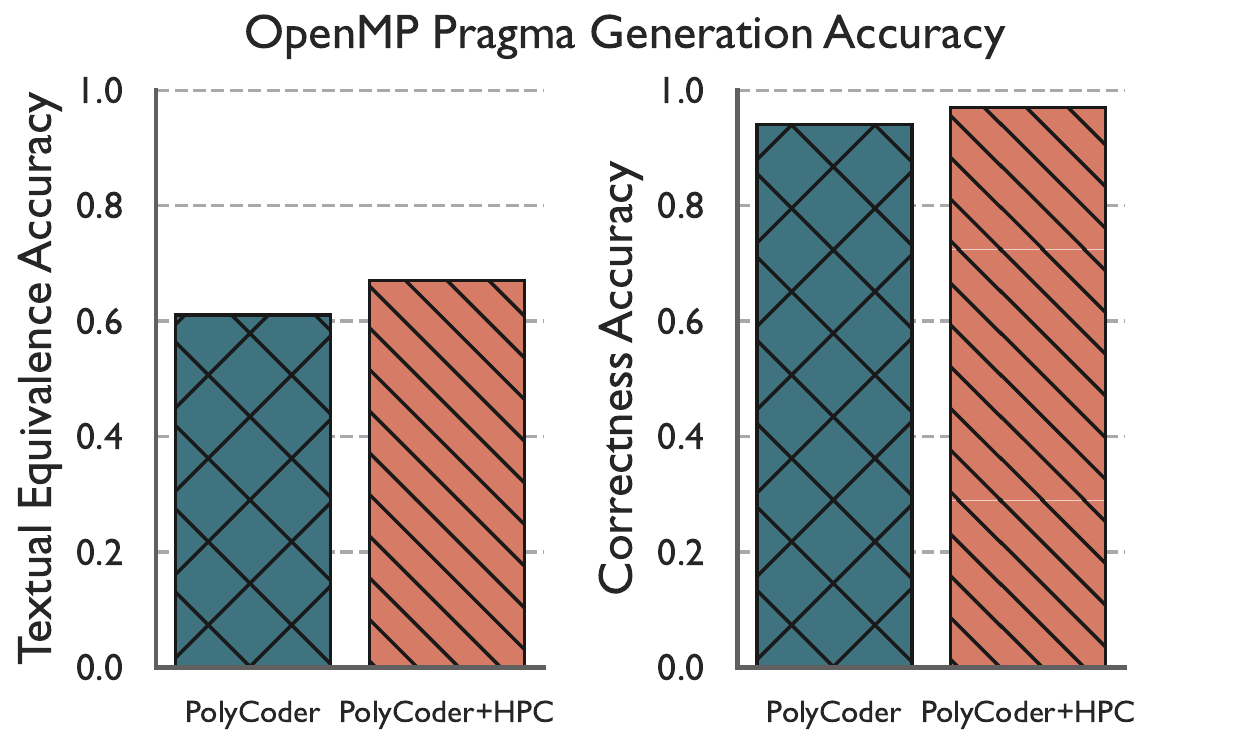}
    \caption{Comparison of models on predicting OpenMP pragmas. The left plot 
    presents accuracy in predicting OpenMP pragmas exactly as they appear in the dataset.
    The right plot shows the accuracy in predicting functionally correct 
    OpenMP pragmas. Higher accuracy is better.}
    \label{fig:openmp-prediction}
\end{figure}

Next, we examine the result from the OpenMP prediction tests described in
Section~\ref{sec:omp-pragmas}. Figure~\ref{fig:openmp-prediction} shows the
results from the OpenMP experiments detailed in Section~\ref{sec:omp-pragmas}.
We see that both models are able to generate functionally correct OpenMP pragmas
with high accuracy (right plot). PolyCoder+HPC is able to do this with 97\% accuracy and
PolyCoder 94\%. The LLMs are exemplary at understanding the
dependencies of the \verb|for| loop and what clauses are required to correctly
parallelize them. We see that the model that has seen large amounts of OpenMP code
performs better.

We can also look at how well the models reproduce the pragmas exactly. This
means all the clauses and variables within those clauses are in the same order
in the dataset and in the output from the model. These results are shown in the
left plot in Figure~\ref{fig:openmp-prediction}. While less meaningful than
functional correctness, it is interesting that the model is able to exactly
reproduce pragmas it has not seen before with relatively high accuracy (67\% and
61\%). This is likely due to certain trends in the construction and ordering of
OpenMP clauses that the LLMs are learning as they train.

\subsection{Relative Performance Prediction}

\begin{figure}[h]
    \centering
    \includegraphics[width=\columnwidth]{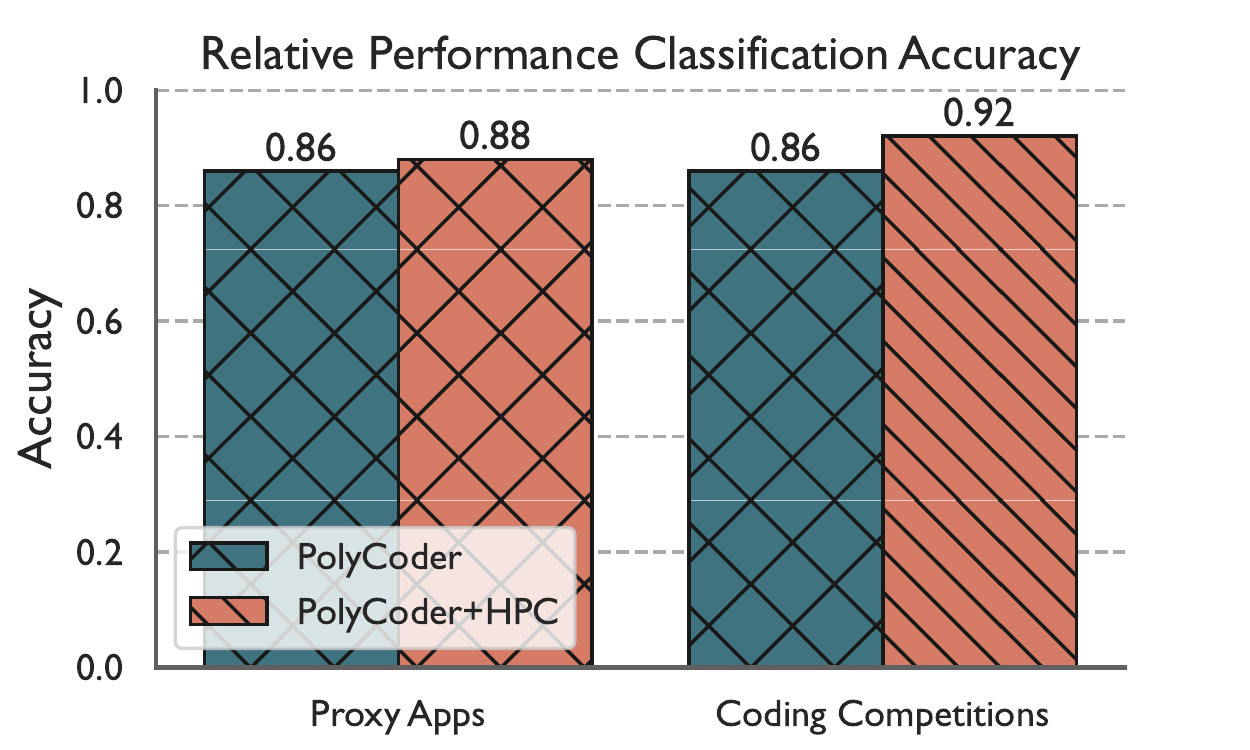}
    \caption{Comparison of models on predicting relative performance of code
    changes. Both models achieve similarly high accuracy. The PolyCoder+HPC
    model performs slightly better on both datasets. Higher accuracy is
    better.}
    \label{fig:rel-perf-prediction}
\end{figure}

Finally, we look at the results from the relative performance prediction tests
described in Section~\ref{sec:rel-perf-prediction}.
Figure~\ref{fig:rel-perf-prediction} shows the results from the relative
performance prediction tests (see Section~\ref{sec:rel-perf-prediction}). Both
models achieve high classification accuracy with PolyCoder+HPC being slightly
better for the two proxy applications at 88\% and PolyCoder at 86\%. This means
that for 88\% of the code changes in the two repositories version control
history PolyCoder+HPC is able to correctly identify if there will be a
performance slowdown. Likewise for the programming competition dataset we see
that PolyCoder+HPC outperforms the PolyCoder baseline with an accuracy of 92\%
vs 86\%. This is a higher accuracy improvement than the proxy applications 
by 4 percentage points. This is likely due to the fact that the programming
competition dataset is larger and PolyCoder+HPC has been trained
on more C/C++ code.

The success of this test demonstrates that the models are able to correlate
their prior language understanding with performance related properties of code.
This means we can leverage LLMs and fine-tuning to model code performance
without the need to collect large amounts data.

\section{Related Work}
\label{sec:related-work}
In this section we detail related work that uses LLMs to study source 
code and work that uses machine learning to model the performance of 
source code.

\subsection{LLMs for Code Generation}
With the explosion in research in transformer models and LLMs there have been a
large number of papers applying these techniques to source code. Most of these
methods have extended GPT-2~\cite{gpt-2}, GPT-3~\cite{gpt-3}, or
BERT~\cite{bert,roberta} models and trained them on code. A notable instance is
Codex~\cite{codex-copilot-short-author}, which is a modification of GPT-3 that
is targeted for source code generation. Following Codex's introduction there
have been several other works that have introduced state-of-the-art large
language models~\cite{li2023starcoder,roziere2023code,wei2023magicoder}. While
some of these are open source, the best, such as GPT-4~\cite{openai2023gpt4},
keep their architecture, weights, and training data closed source and only
inference is available via a paid API.

A large amount of this recent research has focused on code generation.
These usually take a mix of code and natural language and learn how to 
meaningfully finish the code.
While seminal works have continued to improve code generation with better 
and bigger models~\cite{codex-copilot-short-author,gpt-3,roberta}, other 
works have explored how to better utilize these tools in software engineering
workflows~\cite{Dderlein2022PilotingCA,Barke2022GroundedCH,Sarkar2022WhatII}.
Some flip code generation around and learn to generate natural language 
code summaries from code 
snippets~\cite{Gu2022AssembleFM,Ahmed2022LearningCS,Haque2022SemanticSM,Ahmad2020ATA}.

These models can even be trained for tasks such bug and malware
detection~\cite{Richter2022CanWL, Kharkar2022LearningTR}. LLMs can also be used
to suggest fixes in these cases rather than just identify problematic code. Many
other previously difficult to automate software development tasks have since
been tackled by applying LLMs~\cite{ml_for_code}. More recently some of these
tasks have included HPC development tasks such as race
detection~\cite{chen2023data} and OpenACC compiler
validation~\cite{munley2023llm4vv}.

\subsection{Machine Learning Applied to Source Code Performance}
However, one important problem in software development that has not received
much research with LLMs is that of performance.
Many of the reasons listed in Section~\ref{sec:introduction} have prevented 
meaningful studies from being accomplished.
Previously approaches used code2vec~\cite{code2vec}, ir2vec~\cite{ir2vec_2020},
or a similar method to first map source code to an embedded space that could 
then be learned on.
These were successfully used for some performance related analytical modeling 
such as OpenCL kernel device placement~\cite{ir2vec_2020}, but never 
leveraged LLMs for a full performance study.

Garg et al.~\cite{Garg2022DeepDevPERFAD} recently introduced DeepDevPERF, which 
is a BART-based~\cite{bart} LLM designed to suggest performance improvements to
arbitrary C\# code.
They overcome the issue of data collection by using code changes from Git 
commits that have performance related keywords in their commit message,
albeit, this dataset is still noisy.
This work is different than that presented in this paper as it suggests 
code transformations rather than learn relative performance. The latter 
being useful in cases where two versions of a code already exist, such as 
with Git commits.
Additionally, our model is trained on real performance data and can be used 
for HPC and parallel code generation tasks.

\section{Conclusion and Future Work}
\label{sec:conclusion}
In this paper, we have demonstrated the fine-tuning of an LLM using HPC code,
and its ability to outperform other LLMs in HPC related tasks such as HPC code
generation and performance modeling.  We have accomplished this by fine-tuning
a model, and showing that it can generate functionally correct HPC code at up
to a 53\% higher pass@k rate and can accurately label \verb|for| loops with
OpenMP pragmas with 97\% success.  We have further demonstrated how this
fine-tuned model can be utilized to study performance properties of source code
with little data.  These results demonstrate the need for and usefulness of
HPC-specific language models. The best model in our experiments, PolyCoder+HPC, we present as
\modelname{}.

In the future, we plan to explore further analyses that can be accomplished using
our language model.  We also plan on exploring how to tune the model to
generate not just correct but performant code.  Additionally, we plan
to investigate how to engineer these innovations into practical tools that can
be easily used by computational scientists and HPC developers to enable them to produce
better code more efficiently.

\section*{Acknowledgment}
This material is based upon work supported in part by the National Science
Foundation under Grant No.~2047120. This work was performed in part under the
auspices of the U.S.~Department of Energy by Lawrence Livermore National
Laboratory under Contract DE-AC52-07NA27344 (LLNL-CONF-844549).

\IEEEtriggeratref{37}
\bibliographystyle{IEEEtran}
\bibliography{bib/pssg,bib/cite}

\end{document}